\documentclass[sigchi, nonacm=true]{acmart}

\AtBeginDocument{%
  \providecommand\BibTeX{{%
    \normalfont B\kern-0.5em{\scshape i\kern-0.25em b}\kern-0.8em\TeX}}}

\copyrightyear{2019}
\acmYear{2019}
\setcopyright{rightsretained}

\acmConference{}{2019}{USA}
\acmDOI{}
\acmISBN{}
\acmBooktitle{}

\begin{document}

\title{Pretrained AI Models: Performativity, Mobility, and Change}

\author{Lav R.\ Varshney}
\affiliation{%
  \institution{Salesforce Research}
  \city{Palo Alto, CA}
  \country{USA}}
\email{lvarshney@salesforce.com}
\orcid{0000-0003-2798-5308}

\author{Nitish Shirish Keskar}
\affiliation{%
  \institution{Salesforce Research}
  \city{Palo Alto, CA}
  \country{USA}}
\email{nkeskar@salesforce.com}

\author{Richard Socher}
\affiliation{%
  \institution{Salesforce Research}
  \city{Palo Alto, CA}
  \country{USA}}
\email{rsocher@salesforce.com}

\begin{abstract}
The paradigm of pretrained deep learning models has recently emerged in artificial intelligence practice, allowing deployment 
in numerous societal settings with limited computational resources, but 
also embedding biases and enabling unintended negative uses.  In this paper, we treat pretrained models as objects of study and discuss the ethical impacts of their sociological position.  We discuss how pretrained models are developed and compared under the common task framework, but that this may make self-regulation inadequate. Further how pretrained models may have a performative effect on society that exacerbates biases.  We then discuss how pretrained models move through actor networks as a kind of computationally immutable mobile, but that users also act as agents of technological change by reinterpreting them via fine-tuning and transfer.  We further discuss how users may use pretrained models in malicious ways, drawing a novel connection between the responsible innovation and user-centered innovation literatures.  We close by discussing how this sociological understanding of pretrained models can inform AI governance frameworks for fairness, accountability, and transparency.
\end{abstract}

% <concept_id>10003456.10003457.10003521.10003524</concept_id>
% <concept_desc>Social and professional topics~History of software</concept_desc>
% <concept_id>10010147.10010257.10010293</concept_id>
% <concept_desc>Computing methodologies~Machine learning approaches</concept_desc>
% <concept_significance>300</concept_significance>

% \ccsdesc[500]{Social and professional topics~History of software}
% \ccsdesc[300]{Computing methodologies~Machine learning approaches}

%%
% \keywords{artificial intelligence, deep learning, science and technology studies}

\maketitle

\section{Introduction}
\label{sec:intro}

Large-scale deep learning models with billions of parameters can now perform a variety of natural language and vision tasks at or above human levels, but require significant computational---and therefore energetic/monetary---resources to train.  As such, the development of these models has largely been carried out by artificial intelligence (AI) researchers in large institutions (especially for-profit companies) \cite{SchwartzDSE2019}, and is out of reach for researchers in smaller institutions/academia and other technically-skilled AI enthusiasts.  For brevity, we will sometimes refer to these social groups as \emph{producers} and \emph{lead users}, following the user-based innovation \cite{Hippel2017} and social construction of technology (SCOT) \cite{KlineP1996} literatures.  

The development and release of \emph{pretrained} deep learning models by producers has recently emerged as a standard paradigm in AI practice, allowing lead users to then fine-tune and transfer them for use in a variety of research and societal settings.  In natural language processing, examples of pretrained models include BERT \cite{DevlinCLT2019}, GPT-2 \cite{RadfordWCLAS2019},
ELMo \cite{Peters_ea2018}, and XLnet \cite{YangDYCSL2019}.  Beyond the models themselves, producers may or may not also release the training dataset, the code implementing the learning rule, or descriptions of the computational infrastructure; this provides varying levels of transparency.  Unfortunately, as we detail in the sequel, pretrained models may embed biases in unknown and immutable ways while also enabling unintended negative uses.

In this paper, we treat pretrained models as objects of study and discuss the impact their sociological position has on fairness, accountability, and transparency in the larger sociotechnical systems in which they are embedded.  We draw on analytical frameworks from science and technology studies (STS).  Without taking a strong normative position, we especially focus on implications for AI governance processes. 

Whether considering nuclear reactions, recombinant DNA technology, or mutant flu strains, much scientific research and innovation can benefit the public but also be diverted to harmful uses.  A typical reaction by scientists performing such dual-use research has been self-regulation and self-imposed moratoriums, yet careful historical study demonstrates the inadequacy of this.  As Kaiser and Moreno argue, ``no matter the field of research, can anyone be expected to step outside the excitement and momentum of their own work to make objective decisions in risky situations?'' \cite{KaiserM2012}.  Here we suggest such momentum may be even stronger when entire research fields are oriented around a quest to achieve a singular objective---\emph{Holy Grail performativity} \cite{Varshney2014} in Austin's sense of concepts being performed in practice \cite{Austin1970, MacKenzie2006}.  The ascendancy of the so-called common task framework in AI \cite{Donoho2019} embodies exactly such performativity, yet the dual-use potential of pretrained models has led to recent attempts at self-imposed limits on open release \cite{RadfordWAACBS2019, Heaven2019}.  The growing \emph{responsible innovation} literature within science and technology studies \cite{StilgoeOM2013} has been discussed in relation to AI by Brundage  \cite{Brundage2016,BrundageG2019}, but these self-regulation actions are seemingly not informed by understanding the position of pretrained models that emerges from treating them as objects of sociological study, cf.~\cite{Brundage_ea2019}.  We will discuss how insights from responsible innovation and broader STS discourse may inform AI governance policies.

The responsible innovation literature has, as far as we can tell \cite{KerrHT2018}, remained unconcerned with user-driven innovation \cite{Hippel2017} and users as agents of technological change \cite{KlineP1996,OudshoornP2003}.  Yet, user innovation is of central importance in AI, where innovation comes not just from producers of pretrained models but also lead users of pretrained models that fine-tune and transfer them to applications outside the control (and often outside the imagination) of the producers.\footnote{Note that in this paper, we do not consider the final consumers of AI inferences, which are another social group altogether.}  When pretrained models are at the consumption junction (in the sense of Cowan \cite{Cowan1987}), they may be reinterpreted in malicious ways.  Taking the case of AI, we will discuss how principles of governance from responsible innovation should be extended to consider the role of users as innovators.  Of particular relevance for this extension is to understand how pretrained models are developed and evolve as they move among actors in the two social groups.  

Von Hippel and colleagues have noted a kind of division of labor in types of innovations pursued by producers and users for scientific instruments \cite{RiggsH1994} and whitewater kayaking \cite{HienerthHJ2014}.  Producers pursue innovations of interest to the entire market, typically along a fixed dimension of merit such as faster, cheaper, or more reliable.  Contrarily, users pursue innovations to do functionally new things without strong concern for the fraction of the market that may be interested, since they are self-rewarded through intrinsic motivations \cite{BaldwinHH2006,Hippel2017}.  A similar division is seen in AI where an initial general-purpose model like BERT \cite{DevlinCLT2019} is developed by an industrial research lab (and improved by another industrial research lab as 
RoBERTa \cite{Liu_ea2019}) whereas academic researchers fine-tune/transfer such a model to have other more specific functionality such as BioBERT \cite{Lee_ea2019} (for medical text mining), ViLBERT \cite{LuBPL2019} (for vision-and-language tasks), and BERTserini \cite{YangXLLTXLL2019} (for question-answering).

Although pretrained models are not diagrams or inscriptions having the possibility of optical consistency in the sense of Latour \cite{Latour1986}, they do move around among actors in the AI community and are a kind of \emph{computationally} immutable mobile where the immutability stems from the computational costs in modification.  Despite the consistency of pretrained models---at the level of individual bits---as they move, they remain \emph{interpretively} flexible for users.  Since these models are largely obtained anonymously from open source repositories rather than from personal instructional interaction (like Feynman diagrams \cite{Kaiser2005}), further interpretive flexibility is maintained. In fact it is this plasticity of pretrained models in the hands of users---who can fine-tune them for transfer to alternative tasks---that has given them their staying power. 

Moving from responsible innovation considerations due to intentional malicious use by users, we also consider unintentional ethical issues such as algorithmic unfairness that may be immutably embedded in pretrained models.  Since these models are often abstracted by users as essentially black boxes with general intelligence ability that can be fine-tuned to transfer to any task, the biases in models and their training data are not considered \cite{SelbstBFVV2019}.  Moreover, due to computational immutability, these biases are fixed. 
More troublingly, as we detail later, the descriptions of the world (including societal biases) embedded in pretrained models have Barnesian performativity \cite{Barnes1983,MacKenzie2006}, in the sense they may act to shape the future evolution of the world.  That is, as noted in classical economic theories of discrimination \cite{CoateL1993,Spence1973,Arrow1973} and recent models of model retraining \cite{MouzannarOS2019, HuC2018, DobbeDGK2018}, populations might become more like what (biased) models predict.  In fact, biases may even be amplified through the fine-tuning and transfer carried out by lead users \cite{ZhaoWYOC2017} in a single stage of evolution. We will further discuss how this understanding of unfairness propagation can inform AI governance.

To summarize, we revisit responsible innovation in the context of AI fairness, accountability, and transparency by characterizing the social position of pretrained models:
\begin{itemize}
    \item Holy Grail performativity in model development due to the common task framework,
    \item Users as innovators and agents of technological change through fine-tuning and transfer, 
    \item Computational immutability but interpretive flexibility of pretrained models as they move among actors, and
    \item Barnesian performativity of pretrained models in terms of the evolution of algorithmic fairness.
\end{itemize}

\section{Self-Regulation and the Holy Grail of Pretrained Model Development}
\label{sec:grail}

In February 2019, OpenAI developed a large-scale unsupervised language model called GPT-2 (Generative Pretrained Transformer 2) to generate several coherent sentences of realistic text by extending any given seed.  This model further simultaneously performs well on a variety of language tasks including question answering, reading comprehension, summarization, and translation \cite{RadfordWCLAS2019}.  We should further note that in general, better pretrained models lead to better performance on fine-tuned or transfer tasks \cite{MahajanGRHPLBM2018,KornblithSL2019}.

Contrary to recent practice in the artificial intelligence community, OpenAI did not release the training data or the learned parameters of their largest neural network model, only smaller ones.  This, due to concerns that large language models may be used to generate deceptive, biased, or abusive language at scale. In describing their decision to limit the transparency of the GPT-2 model, the producers described several positive and negative uses, which we quote here \cite{RadfordWAACBS2019}:
\begin{quotation}
[+] AI writing assistants, more capable dialogue agents, unsupervised translation between languages, and better speech recognition systems 

\noindent [--] Generate misleading news articles, impersonate others online, automate the production of abusive or faked content to post on social media, automate the production of spam/phishing content
\end{quotation}
As seen, the producers themselves did not specify too many functionally novel uses.  Yet, lead users quickly transferred the model to multifarious settings; a positive example using a smaller version of GPT-2 that OpenAI did release, Deep TabNine is a software programming productivity tool\footnote{https://tabnine.com/blog/deep} to predict the next chunk of code, fine-tuned on open source files from GitHub capturing numerous programming languages.

Notwithstanding numerous arguments against GPT-2 actually posing a societal threat \cite{Heaven2019}, the self-regulation practiced by OpenAI is rather limited. It is not a self-moratorium but only a limitation on distributing detailed results (the pretrained model itself).  Indeed, a student with significant computational resources provided by Google \cite{Leahy2019} purportedly reproduced OpenAI's GPT-2 model, though he also did not release his model for verification \cite{Leahy2019b} citing similar concerns of malicious uses, especially with respect to setting social norms for future release of dual-use AI technology.

As noted in Section~\ref{sec:intro}, several cases in the history of science have shown that self-moratoriums are ineffective, to say nothing of limited self-regulation that does not militate the pursuit of technological progress and may even encourage it.  After all, knowing that something can be done is often a greater spur for future innovation than a detailed description of how it was done.

In the next subsections, we describe certain social norms among producers of pretrained models and then discuss why these norms render self-regulation inadequate.

\subsection{Pursuing Holy Grails}
In building engineering systems---whether physical systems like engines or informational ones like AI---benchmarking performance to understand how well one is doing is often cast as important. To do so, scientists try to both establish clear metrics of performance (often measured in standardized units) and have useful points of comparison. In this vein, the Scottish engineer James Watt developed the concept of horsepower to benchmark the output of steam engines by comparing to the power of draft horses. Indeed, comparing performance of new technologies with either humans or animals that have similar abilities is a typical strategy.

In AI, the Turing test has been proposed as a way to measure a machine's ability to exhibit intelligent behavior by making a binary comparison to people. A machine is said to be intelligent if it exhibits behavior equivalent to, or indistinguishable from, that of a human. This is to be tested through a conversation with human judges \cite{Turing1950}. There are well-known limitations of the Turing test in terms of gaming, cheating, and operational difficulty. It is also largely focused on language ability at the neglect of other facets of intelligence such as perception or creativity. As such, several alternatives have recently been proposed \cite{MarcusRV2016}.

\begin{figure}
    \centering
    \includegraphics[width=2.8in]{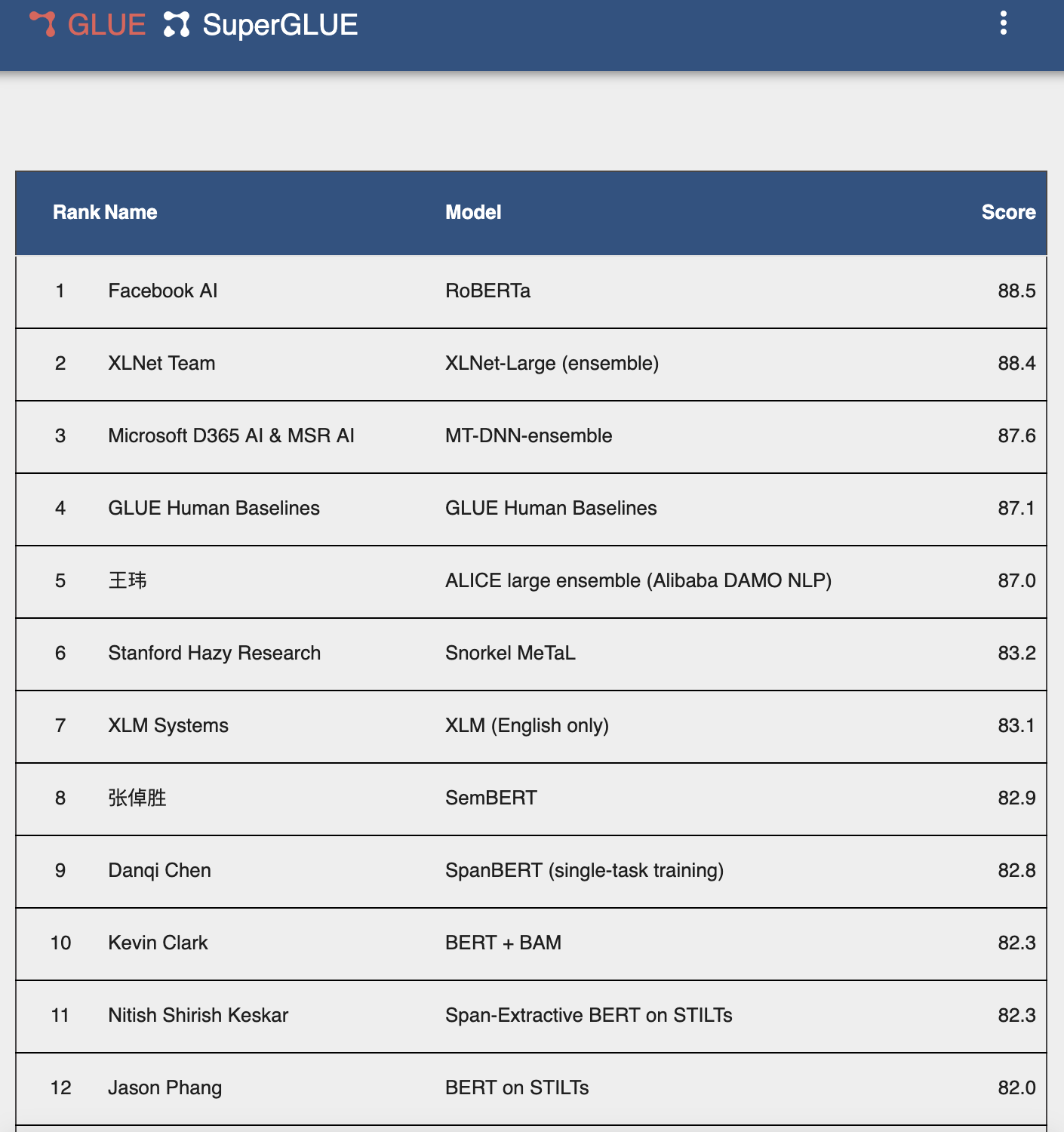}
    \caption{The GLUE leaderboard for general performance on natural language tasks (14 Aug. 2019) \cite{WangSMHLB2019}. Notice the graded performance on a single dimension,  comparing to human performance (\#4).}
    \label{fig:leaderboard}
\end{figure}

One basic property of these new test proposals is scoring intelligence in a graded manner, rather than just all-or-nothing, cf.~\cite{ClarkE2016,AdamsBC2016}. This provides a refined characterization of system performance on a quantitative scale.  Yet, these approaches still essentially use human performance as a benchmark for comparison, even though there is much variation in human intelligence not only within populations but even across the historical record \cite{Neisser1997} (and may therefore not be absolute milestones, contrary to \cite{Shieber2016}).  A typical AI leaderboard oriented as the pursuit of human performance is shown in Figure~\ref{fig:leaderboard}, here measuring performance with standard evaluation data (an idea we will return to).
 
An alternative to judging performance relative to animals or humans is to establish fundamental theoretical limits. Whether considering the Carnot limit on the efficiency of engines \cite{Carnot1824} or the Shannon limit on reliable communication in the presence of noise \cite{Shannon1948}, engineering systems theories establish what is possible and what is impossible. The boundary between the two is what is optimal. Thus, such limit theorems provide absolute standards by which performance may be measured. If a communication scheme operates within $0.0045$ dB of the Shannon limit, this is nearly as good as can be and is independent of how efficiently human communication operates in noise.  Once an absolute scale anchored on fundamental limits is established, human performance can also be fixed as a statistical distribution on that spectrum.\footnote{There are some settings where animals perform intelligent behavior nearly at the fundamental limits \cite{Alexander1996}. Examples include great tits (\emph{Parus major}) nearly achieving optimal performance in feeding strategy, as described by two-armed bandit exploration/exploitation tradeoffs; moose (\emph{Alces alces}) essentially achieving the optimal diet in a Michigan national park, as given by the solution of a linear program; and lions (\emph{Panthera leo}) having hunting behavior over time that matches the solution to a dynamic programming optimization. In all examples, performance of particular behavioral strategies taken by given organisms has been compared to the best performance possible by any strategy.}  

There are several AI settings for which \emph{non-constructive} fundamental limits are known: by non-constructive we mean that although the limits can be computed, strategies that actually achieve those limits are not known.  Examples include flying without crashing \cite{KaramanF2012}, combinatorial creativity \cite{Varshney2019}, communicating with aliens \cite{Misra2010}, and reconstructing the tree of life \cite{SteelS2002}.  In data-driven areas of AI, however, it is often not possible to define fully closed deductive systems \cite{Netz1999} in which to reason about fundamental limits (even if in principle, Bayes risk is a fundamental limit).  To emulate the kind of abstraction achieved in closed deductive systems, the common task framework has emerged as a prevailing paradigm for AI model development \cite{Donoho2019}.  The idea is to pursue best task performance on a fixed dataset, split into training and testing portions.  Figure~\ref{fig:leaderboard} shows the common task framework in action where the standardized GLUE dataset \cite{WangSMHLB2019} is used to assess performance of different AI models on a standard task set.  
As seen, producers aim to develop AI models that perform better than humans and each other, and the top results are common pretrained models such as RoBERTa and XLNet.

Moreover, because ideals are data- and task-specific, there can be a progression of goals within the common task framework, different than information-theoretic or thermodynamic limits which are fixed by the closed deductive system.  For example, the header of Figure~\ref{fig:leaderboard} indicates that SuperGLUE has been developed as a successor to GLUE.

\subsection{Inadequacy of Self-Governance}

Drawing on the historical case of coding theory being organized as a quest to achieve information-theoretic limits, Varshney had argued that closed universes of deductive discourse and fundamental limits within them lead to Holy Grail performativity \cite{Varshney2014}.  That is, introducing the concept of a \emph{limiting ideal} is performative: the use in practice of a theoretical concept orients research and innovation more towards that theoretical concept. 

As Pierce described \cite{Pierce1965}, again about coding as a quest to achieve information-theoretic limits, ``it may be true that communications theorists could have devised error-correcting schemes even if they never knew of the limit theorems of information theory, but it is doubtful that they would have tried so hard and so well without limit theorems with which to compare their results (and occasionally to goad themselves).''  This strongly captures the central thesis of goal-setting theory, a well-established theory of motivation in psychology \cite{LockeL2002}.  The idea is that the most effective performance seems to result when goals are
specific and challenging.  Further, psychological momentum in pursuing a set goal is difficult to attenuate. Indeed, in Holy Grail performative settings where entire social groups are pursuing the same specific goals, this behavioral momentum is strengthened by social comparison (as facilitated by leaderboards).  When the goals also evolve to become more difficult, this allows actors in the social group to ``level up'', yielding greater motivation.  A side effect of goal setting, however, may be a narrow focus that neglects non-goal areas \cite{OrdonezSGB2009}.

These behavioral factors are redolent of Kaiser and Moreno's claim that innovators cannot be expected to step outside the momentum of their work to self-regulate.  Developing AI models within the common task framework has Holy Grail performative social norms, much more so than, say, DNA recombination where innovators had disparate functional goals.  As such, self-regulation is especially inadequate and alternative governance approaches developed within the responsible innovation literature should be considered.

\section{Users as Fine-Tuners of Pretrained Models}
\label{sec:users}

As we have seen, the culture of AI model producers is very much Holy Grail performative, pursuing innovation along a dimension of merit like the GLUE score in Figure~\ref{fig:leaderboard}.  In this section, we turn to the social group of lead users, who are concerned with functionally new applications of AI models \cite{Hippel2017}.  As Cowan has argued \cite{Cowan1987}, analysis focused on users allows for the possibility of unintended consequences, ``without which no sociological or historical explanation should be taken seriously''.  Yet, the responsible innovation literature has remained unconcerned with user-driven innovation, cf.~\cite{StilgoeOM2013}. 

Within AI  governance, too, the distinct role of innovative users seems to be unconsidered, see e.g.\ a recent survey on AI ethics frameworks that does not consider the social group of users \cite{Hagendorff2019}.

\subsection{Users as AI Innovators}
User innovation is of central importance in AI, where innovative lead users of pretrained models fine-tune and transfer them to functionally new applications, often far beyond what producers may have imagined.  Although we do not discuss it further here, a closely related setting is multi-tenant cloud provision of AI models where the model creator does not have access to the data or application scenario for which the customer is deploying the model.

\begin{figure}
    \centering
    \includegraphics[trim=100 160 100 250, clip, width=4in]{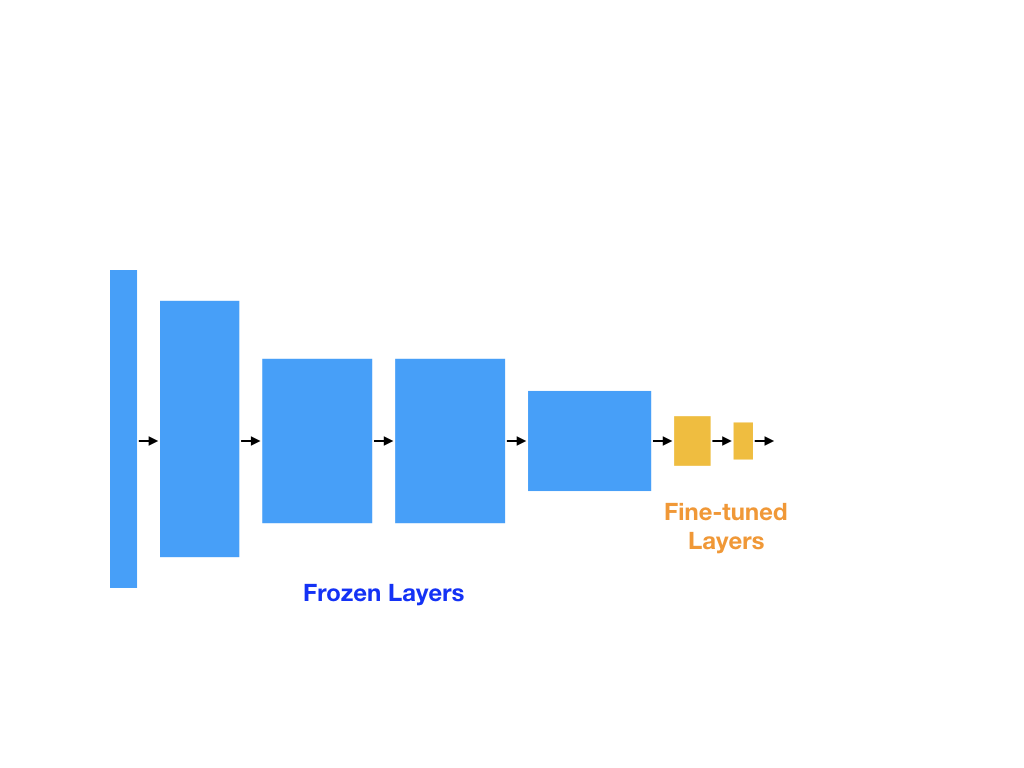}
    \caption{In transferring a pretrained neural network model to a functionally new task, the early layers may be frozen as is, and the last few layers retrained using new data for the new task.}
    \label{fig:finetuning}
\end{figure}

Before proceeding, let us briefly describe the technological approach used for transferring a neural network model developed for one task to work on a second task by fine-tuning.  In deep neural networks---taking feedforward networks such as multilayer perceptrons or convolutional neural networks as examples shown in Figure~\ref{fig:finetuning}---it has been found that early layers of models produce features that capture general attributes of the training dataset whereas later layers of the model capture properties of the task it is trained on.  As such, one approach for inductive transfer of a model for one task (the pretrained model) to become a model for a different task (the fine-tuned model) is to freeze the early layers from the pretrained model and retrain the last couple layers using a new data set and a new task.  This is computationally much easier than training a new model from scratch: since most learned parameters are taken straight from the pretrained model, a much smaller number of parameters must be learned. In essence, this works by beneficially narrowing the scope of possible models for the new task.  The basic idea is depicted in Figure~\ref{fig:finetuning}, using the now-standard diagrammatic style for neural network architectures.

Putting nuclear technology to new uses requires large, expensive facilities and using recombinant DNA technology requires specialized reagents, but fine-tuning and transfer of AI models does not require either.  Fine-tuning AI models is feasible for a large social group of lead users.  Even though producers of pretrained models may have a particular meaning in mind, they do not control how these artifacts are used once in the hands of users.  As noted in the social construction of technology literature, ``users precisely as users can embed new meanings into the technology'' \cite{KlineP1996}.  Indeed, it is well-established in the user-driven innovation literature \cite{Hippel2017} that lead users come up with numerous functionally new applications. 

In the early twentieth century, Ford had built the Model T with the singular interpretation as a passenger vehicle, but rural American users put it to use as a power source for washing machines, butter churns, cream separators, corn shellers, water pumps, hay balers, fodder and ensilage cutters, wood saws, hay and grain hoists, cider presses, and corn grinders, as well as in mobile form as a snowmobile, tractor, and agricultural transport vehicle. This  interpretive flexibility of users later pushed Ford Motor Company itself to create modification kits for the Model T \cite{KlineP1996}.  Although perhaps not quite as general-purpose technology as a car engine, pretrained AI models have also been put to use in numerous settings.  Taking the example of the BERT language model \cite{DevlinCLT2019}, it has been used for clinical medicine (clinicalBERT), scientific research (SciBERT), story generation (TransBERT), and intellectual property law (PatentBERT), among many other language task settings just within a year of its release.

Besides these functionally new innovations that are putatively societally beneficial, lead users of AI models have also reinterpreted them perniciously to innovate in societally harmful ways.  A typical example is DeepNude, an app that removes clothing from the images of women, making them look realistically nude \cite{Cole2019} and is based on the previous pix2pix image transfer model \cite{IsolaZZE2017}.  Katelyn Bowden, founder and CEO of revenge porn activism organization Badass, was quoted as saying ``Now anyone could find themselves a victim of revenge porn, without ever having taken a nude photo. This tech should not be available to the public'' \cite{Cole2019}.

Users act as agents of technological change \cite{OudshoornP2003} not only in changing the interpretation of artifacts as we have described, but also in shaping the future design of artifacts themselves \cite{KlineP1996}.  The relationships among social groups both constrain and enable the design and usage of technology, and the social groups in turn get shaped in designing/using the technology.  

The design and use of skateboards took place as community-based innovation, with significant back-and-forth among producers and users \cite{Shah2006}. AI models are also developed largely within a tightly knit community.  Academically-inclined producers and users publish papers in the same scholarly conferences; individual innovators may be users during their training at universities and then become producers when they join large companies (while also doing industrial internships in between).  Senior researchers may frequently move between universities and industry, even simultaneously having dual appointments in both \cite{Kwok2019}.  Moreover, notwithstanding Section~\ref{sec:grail}, deep learning is largely an open source community which further enables interactions among actors.

In general, pretrained models that perform better on their benchmarks also perform better after fine-tuning on transfer tasks \cite{MahajanGRHPLBM2018,KornblithSL2019}.  Yet, the strong interaction of users with producers through community links has led to pretrained models that are specifically designed to be better at inductive transfer to other tasks \cite{McCannBXS2017,HowardR2018}.  As a notable example, consider SpanBERT (not a fine-tuning of BERT, but a new pretrained model inspired by BERT) designed to be better at inductive transfer to new language tasks \cite{JoshiCLWZL2019}; its developers are primarily from Facebook but also have participants from academia.  The basic idea is to design the pretrained model to better represent and predict spans of text, which arise in several functionally novel language tasks; BERT was concerned with individual words rather than spans of words.  As a variation, pretrained models can specifically be trained for multiple tasks simultaneously \cite{McCannKXS2018}, aiming to generalize well to any task.

We have seen that the flexible interpretation and needs of lead users of pretrained models both influence the design of future pretrained models and lead to functionally new innovations through fine-tuning.

\subsection{Inadequacy of Producer-Focused Governance}
Although producers and lead users are coupled within the AI community, there is a division of labor between the two social groups, which imply distinct considerations for AI governance.  As such, a focus solely on governance for producers would neglect the network of social relations among actors in the AI ecosystem, and the nature of accountability propagating through the actor network \cite{Floridi2016,Cowan1987}. 

Pretrained models at the consumption junction, as Cowan describes it \cite{Cowan1987}, may be interpreted in both beneficent and maleficent ways and therefore yield both putatively positive and negative unintended consequences.  Prima facie, beneficence and non-maleficence are desirable, but these must be balanced in AI governance, as embedded in the interaction network of producers and users.

The case of pretrained AI models suggests that responsible innovation should be expanded to include the role of users.

\section{Pretrained Models as (Im)mutable Mobiles}
\label{sec:mobiles}

We have seen in the previous section that users interpret pretrained AI models in various ways, and transfer them to numerous functionally new uses through fine-tuning.  In this section, we look more at how the models themselves move through the relevant social groups and how an understanding of such information spreading may inform AI governance.  A recent survey indicates that the spreading dynamics, fine-tuning, and recombination of AI models do not enter into existing AI governance frameworks \cite{Hagendorff2019}, which instead focus only on initial development and release.

Just as in Section~\ref{sec:users}, this omission suggests the value in expanding the scope of responsible AI innovation from a static focus on release to considering network dynamics. 

\subsection{Pretrained Models in Action}

Pretrained models are mathematical objects that specify particular neural network architectures and learned synaptic weights: they are functional and can be used directly to perform inference when deployed as AI services or as part of larger AI services in sociotechnical systems, cf.~\cite{Arnold_ea2018,Varshney2016}.  Although they are formalisms, they are not abstractions (under common definitions), but the thing itself.  Abstractions such as neural network architecture diagrams of the type in Figure~\ref{fig:finetuning} also move around---with Latourian optical consistency \cite{Latour1986}---among actors in the AI social network, but we focus on the pretrained AI models themselves.\footnote{Neural network architecture search and hyperparameter tuning is even more computationally intensive than training single neural networks \cite{StrubellGM2019} and so architecture diagrams may take a similar sociological position as pretrained models.}  

Pretrained models move around with not just optical consistency, but mathematically precise identicality.  Indeed, the raison d'\^{e}tre of \emph{pretraining} models is to move with no change, due to the computational cost in training large AI models.  In this sense, they are \emph{computationally} immutable mobiles; yet, as we saw in Section~\ref{sec:users}, they are \emph{interpretively} flexible.
They are essentially physically immutable like car engines, which are difficult to modify without specialized equipment, rather than mutable like paper tools that are inherently plastic \cite{Kaiser2005}.

\begin{figure}
    \centering
    \includegraphics[width=3.4in]{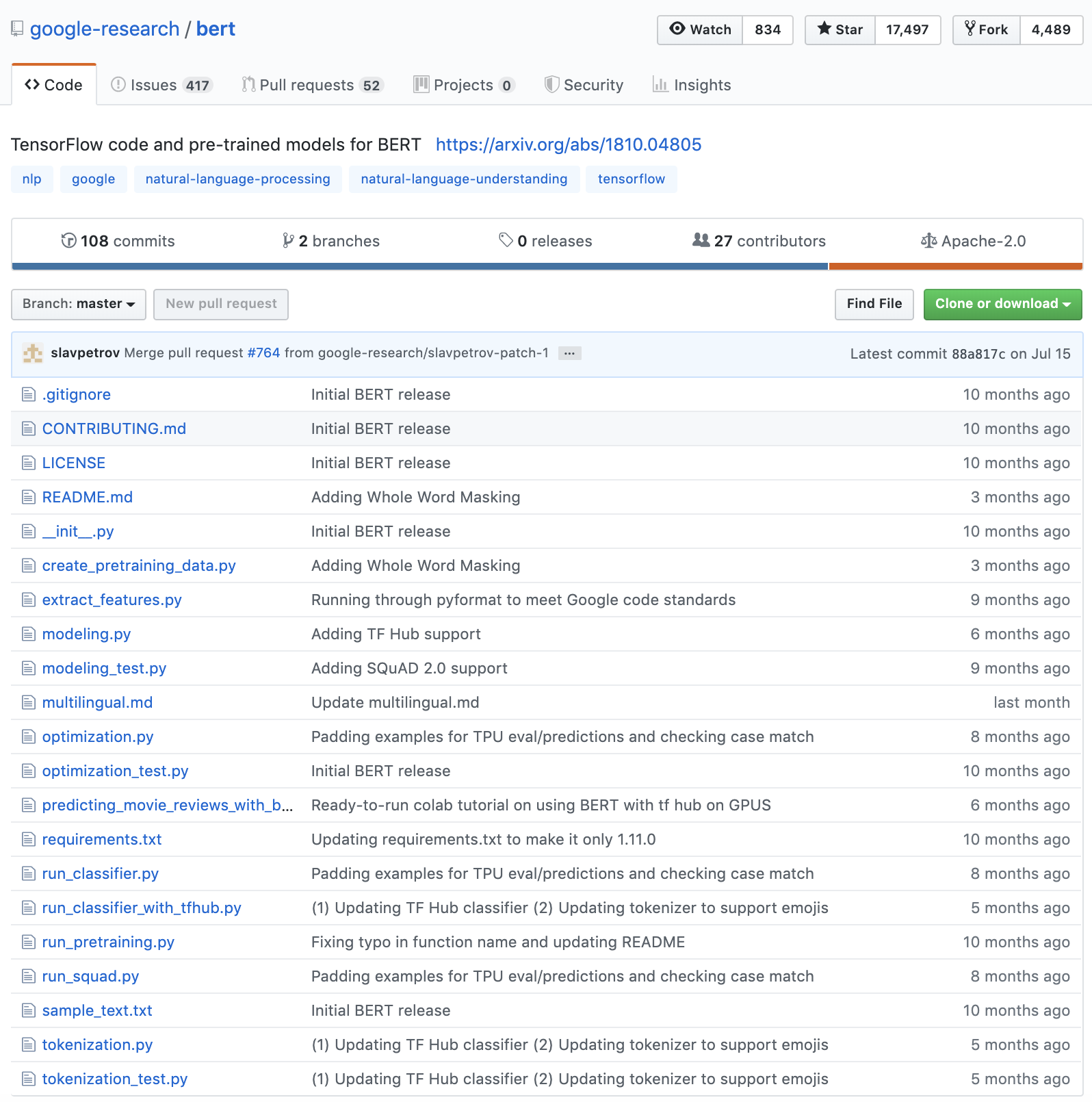}
    \caption{Google Research's BERT GitHub repository, which allows anonymous download of the pretrained model, training code, and documentation, as well as forking and other version management operations (19 Aug. 2019).}
    \label{fig:github}
\end{figure}

Although there are local, personal instructional interactions (including academic training relationships) among actors in AI, the primary way pretrained models are disseminated is through postings to open source repositories such as GitHub, see e.g.\ Figure~\ref{fig:github}. Insights into design ideas and detailed performance characterization are disseminated through preprint servers such as arXiv, together with more informal explanations as blog posts, which may further spread through social media such as Twitter.  In such a technology-mediated open source community \cite{Benkler2006}, models circulate widely from their original point of dispersion.  One can see more than 4000 forks of the BERT model in Figure~\ref{fig:github} by a wide variety of users, to say nothing of downloads that were then fine-tuned. 

Despite limited institutional gate keepers in open source settings (like journal editors, as in some branches of science), cultural norms do lead to a kind of file drawer problem \cite{EvansF2011} where only effective (with respect to producers' benchmarks) models and approaches are put into circulation by producers.  Ineffective ideas only spread by local instructional interaction through personal contact.

In addition to models, datasets, and code spreading, there have been several suggestions to create model cards for pretrained models \cite{MitchellWZBVHSRG2019} or fact sheets for larger compositions of models as AI services \cite{Arnold_ea2018} that move along with them.  Like nutrition labels for food or parts sheets for electronics, they are meant to be documentation listing performance characterization, contexts for intended usage, as well as properties such as safety (including fairness/explainability), security, and provenance.  By listing contexts for intended usage, such documentation is meant to avoid the portability trap arising from abstraction \cite{SelbstBFVV2019}, as pretrained models move around.  As far as we know, such documentation approaches have not been put into widescale practice.

Although Mitchell et al.\ \cite{MitchellWZBVHSRG2019} suggest model cards may inform users on ``different options for fine-tuning, model combination,'' such documentation does not capture how performance, appropriateness, or safety properties may change under fine-tuning and transfer to other tasks, even though this is a primary mode of use.  Moreover, model cards or fact sheets do not include an \emph{expiration date} for validity in an ever-changing world.   

Arnold et al.\ \cite{Arnold_ea2018} argue that ``systems composed of safe components may be
unsafe and, conversely, it may be possible to
build safe systems out of unsafe components,'' and therefore focus on specific larger AI services composed of AI models.  The danger of compositions may be especially pernicious when AI models are sociotechnically coupled in complex and tight ways  \cite{Perrow1984}. Fact sheets do not consider recombinations of models or novel combinations of models.  Indeed, there is as yet, no compositional calculus for the properties of AI models, like there is in cryptography \cite{LiaoHM2019}; unlike other part sheets \cite{CantonLE2008}, proposed AI documentation does not even indicate how to put pieces from a library together to build more complicated AI services (in sociotechnical systems \cite{Varshney2016}).

To summarize, AI models are computationally immutable but interpretively flexible.  They become dissociated from context as they move around---despite attempts to counter this using detailed documentation.  This is especially the case since dissemination is largely technology-mediated and disconnected from personal interactions.

\subsection{Inadequacy of Static Governance}
Once AI models are developed, they move around.  Indeed, much of the action is in this spreading and reinterpretation.  Since there is a decentralized, technology-mediated network of dissemination and change, there are no Latourian ``centers of calculation'' that maintain their scientific prominence and authority by having people continually return to them from hinterlands.  Prominence within the common task framework may come directly from good performance on specifically stated criteria  (achieving which often requires significant resources).
As such, AI governance that only considers existing centers of production and their initial act of dissemination will be inadequate.

Moreover, the fine-tuning and combining of AI models that happens as they move is not mere bricolage, but is governed by the interpretive flexibility that users have for AI models.  An understanding of the changing and combining dynamics of AI models is essential to effective AI governance.  

The case of pretrained AI models suggests that responsible innovation should be expanded to consider the mechanisms and dynamics of spreading throughout the actor network.

\section{Barnesian Performativity of Unfair Pretrained Models}
\label{sec:barnes}

As AI models move through actor networks, become reinterpreted, and are fine-tuned to transfer to contexts previously unimagined, they remain computationally immutable.  One particular property of pretrained AI models and services that model cards and fact sheets aim to capture is fairness, which necessarily also remains computationally fixed.   

When fine-tuned or composed into larger AI services, pretrained models are often interpreted as abstract black boxes of intelligence dissociated from context, much like grain is abstracted when put into a grain elevator, dissociated from its source \cite{Cronon1991}, or gamete cells are abstracted as reagents when carrying out long-term freezing, dissociated from space and time \cite{Landecker2007}. For example, in describing PatentBERT, Lee and Hsiang \cite{LeeH2019} only say:  
\begin{quotation}
In this work, we leverage the released BERT-Base
pre-trained model (Uncased: 12-layer, 768-
hidden, 12-heads, 110M parameters) \ldots Our implementation follows the fine-tuning
example released in the BERT project \ldots We intentionally keep the code
change as minimal as possible
\end{quotation} 
and never discuss any further properties of the BERT model or the dataset it was trained on.  

Cast as black boxes, the internal properties of pretrained models are not of central interest to many lead users.  Even if there were model cards that specifically call out properties such as fairness and these are brought to the attention of users, these characterizations may fall into a formalism trap of abstraction, since summary statistics would not capture e.g.\ contextual or contestable aspects of fairness \cite{SelbstBFVV2019}.

Yet, there is (appropriately defined) unfairness along many socially observable dimensions embedded within pretrained models \cite{MayWBBR2019}.  Despite no animus---only apathy---on the part of actors in the community, this implies unfairness in pretrained models can spread widely.  Moreover, unfairness in AI models can actually exacerbate unfairness in society itself through a kind of Barnesian performativity, as we describe next.  Recall that Barnesian performativity is the effect that using a model in practice makes a societal process more like its depiction by that model \cite{MacKenzie2006}.  Controlling such feedback may require a feedback-based strategy.

Indeed, these kinds of unfairness dynamics for the case of AI suggest the need to expand responsible innovation to consider feedback-based governance.

\subsection{Amplifying Bias in Pretrained Models}

Algorithmic unfairness may be immutably embedded in pretrained models, and further this unfairness may not be evident to users as they often abstract pretrained models as black boxes \cite{SelbstBFVV2019}, dissociated and decontextualized from the training data used to develop them.  When users fine-tune pretrained models to transfer for alternative tasks, recent empirical analyses suggest that they may in fact amplify biases in the original model \cite{ZhaoWYOC2017}.  From a societal perspective, this is similar to adverse drug events from off-label prescribing of drugs \cite{EgualeBVWBJT2016}, where side effects may be amplified when transferring a drug to a clinical setting for which it was not initially developed or tested.

Of greater concern, however, is that societies can perform models, exacerbating the societal bias that was originally present in the training data.  
AI models do not stand outside of society;  rather they are part of the infrastructure of modern society \cite{KleinbergJMS2019}.  Therefore AI algorithms do not just passively capture the properties of society, but in fact shape their evolution as intrinsic parts of societal processes.  

As noted by MacKenzie in his study of financial models and markets \cite{MacKenzie2006}, ``the sociologist Barry Barnes has emphasized the central role in social life of self-validating feedback loops.''  As such, he refers to the form of performativity where the use of a model make the model ``more true'' as \emph{Barnesian}, a term we also adopt.  We observe that the use of a biased AI model makes a difference and may significantly alter society to conform more to the model, a self-fulfilling prophecy \cite{Merton1948}.

The basic mechanisms by which a biased AI model can be Barnesian performative is well-understood in economic theories of discrimination \cite{CoateL1993,Spence1973,Arrow1973,KnowlesPT2001}.  Let us describe the two primary dynamic mechanisms in the context of human resource management, where AI models have been used for many years, e.g.\ \cite{MehtaPSVV2013}. First, a worker in a disadvantaged group may fail to invest in her human capital if she knows the employer's AI model is unlikely to suggest she be promoted.  Second, an employer itself may invest less (e.g.\ for training) in a worker from a disadvantaged group if an AI model indicates that the worker will not benefit.  This leads to a self-fulfilling prophecy when new training data is used to update models that capture this under investment by disadvantaged subpopulations.

In the AI context, recent mathematical models of AI model retraining \cite{MouzannarOS2019, HuC2018} capture this dynamic phenomenon of populations  becoming more like what (biased) models predict, a kind of positive feedback. 

\subsection{Inadequacy of Dead-Reckoned Governance}

In control theory, there are two main approaches: feedforward and feedback. Under feedforward control, a system responds to a control signal in a predefined way, whereas under feedback control, the system adjusts the control signal based on how the plant reacts.  In navigation, feedforward is called \emph{dead reckoning} and requires advanced calculation of the exact direction, magnitude, and timing of all actions.  This is nearly impossible to implement for complex systems whose dynamics are uncertain.  

As we argued, the unfairness of pretrained models may amplify as time progresses and as they are transferred to other tasks---a kind of positive feedback.  Given these complex dynamics, AI governance may take inspiration from control theory, which suggests the use of feedback control either alone or in combination with feedforward control.  In particular, positive feedback can be reduced by feedforward damping supplemented by adding negative feedback.

In law and economics, both regulation and litigation are used to mitigate market failures; one dimension of distinction is that regulation is \emph{ex ante} whereas litigation is \emph{ex post}. Strong \emph{ex ante} approaches are often inspired by the precautionary principle \cite{Sapolsky1990}.  In many ways, \emph{ex ante} governance is analogous to feedforward control whereas \emph{ex post} governance is analogous to feedback control. Note that regulatory approaches can be \emph{ex post}; for example the Food and Drug Administration performs postmarket surveillance of drug safety and issues recall notices when a drug is found unsafe.

When considering unintentional ethical issues such as algorithmic unfairness rather than intentional malicious use by users as in previous sections, we still find that responsible innovation should be expanded with a network-centric viewpoint and further allow the possibility of \emph{ex post} governance based on feedback, rather than just \emph{ex ante} governance. 

\section{Responsible Innovation in AI}
\label{sec:RI_AI}

Discourse in science and technology ethics, and responsible innovation in particular, have put forth general principles and frameworks for thinking about technology governance.  Stilgoe et al.\ \cite{StilgoeOM2013} suggest that ``responsible innovation means taking care of the future through collective stewardship of science and innovation in the present.''  This essentially involves asking what kind of future is desired and then asking what kinds of actions should be taken, given there is much uncertainty about the future.  In this approach, ethical governance moves from consequentialism to a question of process.

Thus far, insights from the responsible innovation literature have played a limited role in AI practice \cite{Brundage2016,BrundageG2019}.  As noted by von Schomberg, definitions in technology governance are usually initially made by using analogies, which serve to normalize the new technology.  As understanding of the technology grows, the force of analogies weakens and distinct governance responses can be made \cite{Schomberg2019}.  Here, making analogies between AI and other potentially dual-use technologies such as nuclear and DNA recombination have allowed us to understand the inadequacy of self-regulation.  Moreover, we will see the analogy also suggests an alternative governance approach.  Distinct approaches, however, may be needed to address the inadequacies of producer-focused, static, and dead-reckoned governance that we have identified through an STS (and especially SCOT)-based analysis of how pretrained models move and change through the actions of distinct social groups. The case of AI suggests that to pursue care for the future, responsible innovation must expand to consider dynamics, feedback, and networks of users.

Drawing on the insight that consequentialist governance premised on formal risk assessment has done little to predict many of the most profound impacts of innovation \cite{StilgoeOM2013}, we take a more expansive viewpoint.  Focusing on process to expand beyond current AI governance approaches (without taking a strong normative stand), we suggest the following possibilities to address the inadequacies discussed in the previous sections.  In doing so, we specifically recognize that the social world acts to fundamentally shape technical development at every level.

{\bf Self-Governance} As detailed in the responsible innovation literature \cite{StilgoeOM2013,Schomberg2019}, contrary to self-governance by innovators, an alternative is deliberative and inclusive governance with broad stakeholder involvement.  This aims to diversify the inputs to and the delivery of governance \cite{CallonLB2009}. A process of inclusion forces consideration of questions of power.  One goal is to achieve a consensus set of norms and governance processes that are based on a broad set of values, standardized across the AI community.  In fact, the Partnership on AI has been pursuing exactly this goal \cite{LeibowiczAE2019}, though this effort may be enhanced by greater understanding of Holy Grail performativity.

{\bf Producer-Focused Governance} We have argued that innovation by lead users into functionally new application areas is a key process in AI, outside the control of pretrained model producers.  Extant discussions of AI governance, however, have focused only on producers.  An alternative is an ethics of co-responsibility, where producers and lead users assume shared responsibility \cite{Floridi2016} for intended and unintended consequences, rather than producers being cast as a kind of moral crumple zone \cite{Elish2019}.  Such a network-centric view recognizes the fact that lead users are agents embedded in a network of social relations that limits and controls the technological choices they are capable of making \cite{Cowan1987}.  More specifically, mechanisms such as codes of conduct and ethical technology review conversations may build greater reflexivity in both users and producers.
    
{\bf Static Governance} Since pretrained models are interpretively flexible as they move, a static view of governance does not capture the dynamics of change and recombination as the models are put to numerous innovative uses.  Moreover, a risk-based assessment does not capture the desire to balance the beneficence and non-maleficence of users.  An alternative possibility is governance built on a compositional calculus for pretrained models, paired with anticipation through technology foresight  that specifically considers their mobility and change.  Note that the responsible innovation literature has developed structured ways of performing technology foresight \cite{KarinenG2010}, but grounding in the (im)mutability of pretrained models would only enhance this process.  
    
{\bf Dead-Reckoned Governance} Whereas fixing an \emph{ex ante} governance approach for anticipated malicious uses may be prudent, it also seems incomplete in the face of complex sociotechnical systems involving AI. Recognizing that most innovations are unexpected and hard to forecast (especially functionally new applications) suggests the need for \emph{ex post} surveillance too, much like ongoing monitoring of drug safety.  Such a feedback-based approach is responsive to the power of innovative technology to create the future.  For fairness specifically, there are even batteries of statistical tests that could be administered as pretrained models move into new applications \cite{Bellamy_ea2018, KnowlesPT2001}, but their deployment may be improved by greater understanding of the self-fulfilling prophecy of Barnesian performativity.  Moreover such feedback-based governance enable social learning.

\section{Conclusion}
Our estimates suggest that the cost to train (not considering architecture search or hyperparameter tuning) XLNet was \$50,000, to train RoBERTa was \$60,000, and to train  GPT-2 was \$250,000.  On the other hand, the cost to fine-tune BERT on the SQuAD dataset is estimated to cost only \$3.  That is, it is at least tens of thousands of times more costly to initially develop a pretrained model than to fine-tune it.  This technological distinction has several social consequences.  We have described here, how large-scale AI models are developed, how they are used, how they move around among agents, and what unfairness properties may be embedded and exacerbated in them as they move.

When closed stoves were developed in the eighteenth century, there were various interpretations about their safety.  Their predecessor technology, open hearths, were also dangerous but their ``dangers were dangers that people had coped with for centuries; the risks of stoves were new and thus potentially more worrisome'' \cite{Cowan1987}.  Such worry (now for AI) often yields a desire for governance, but emerging technologies typically fall into an institutional void, where there are few agreed upon governance structures \cite{Hajer2003} and analogies to old technologies may be inadequate.  Here we have argued that analyzing the sociological position of pretrained AI models suggests expanding responsible innovation to several new factors that may yield more responsive and effective AI governance.

%\begin{acks}
%Omitted for double blind review.
%\end{acks}

\bibliographystyle{ACM-Reference-Format}
\bibliography{ethics_socialgood}

\end{document}